\documentclass[12pt]{article}

\newcommand{\be}{\begin{equation}}

\newcommand{\lll}{\langle}
\newcommand{\rrr}{\rangle}
\newcommand{\ee}{\end{equation}}
\title{\bf
Nonperturbative QCD vacuum and colour superconductivity. }
\author{
N.O.Agasian\thanks{e-mail: agasyan@vxitep.itep.ru}, 
B.O.Kerbikov\thanks{e-mail: borisk@heron.itep.ru}, 
V.I.Shevchenko\thanks{e-mail: shevchen@heron.itep.ru} \\
{\it Institute for Theoretical and Experimental Physics} \\
{\it 117218, B.Cheremushkinskaya 25, Moscow, Russia}}
\date{}
\begin{document}
\maketitle
\vspace{1cm}
{\centerline {\bf Abstract}}
We discuss the possibility of existence
of colour superconducting state in real QCD
vacuum with nonzero $\lll {\alpha}_{s}GG\rrr$. 
We argue, that nonperturbative gluonic fields
might play a crucial role in colour
superconductivity scenario. 

\newpage

\section{Introduction}

The behaviour of QCD at high density has become recently
a compelling subject due to (re)discovery of colour superconductivity
\cite{awr1,rssv}. The essence of the phenomenon is the formation
of the BCS-type diquark condensate at densities exceeding by a
factor ($2\div 3$) the normal nuclear density \cite{br}. 
Colour superconductivity
has been studied within different versions of the Nambu--Jona-Lasinio --
type
model \cite{awr1,awr2,br} 
or the instanton model \cite{rssv,cd}. To our 
knowledge, however, no attention has been paid to the fact, that
as compared to real QCD, both approaches miss important nonperturbative  
gluonic content of the theory. 
The NJL model contains the gluon
degrees of freedom in a very implicit way: it is argued,
that high-frequency mode (one-gluon exchange)
after being integrated out gives rise to the effective
four quark interaction (see, e.g. \cite{bbr}). The instanton model
deals only with specific field configurations -- instantons and
antiinstantons. The NJL model with gluon condensate
included has been considered in \cite{ev}
while confining background superimposed on instantons
has been treated in \cite{s}

The role played by the
gluon degrees of freedom in the problem under consideration
is essentially twofold. First, it is assumed, that they are
responsible for producing the quark--quark attractive
interaction leading at small enough temperatures 
to the Cooper pairing (thus gluons play the role of 
phonons over ion lattice speaking  
in condensed matter terms).  On the other hand, the 
vacuum gluon field fluctuations
should be affected by colour superconducting state itself in a way
analogous to the Meissner effect in ordinary superconductor.
The crucial point is which force will win, i.e.
whether superconductivity
will survive or will be destroyed by gluonic fields as it 
happens in the standard BCS theory in the presence of 
strong enough external magnetic field. 

The studies of idealized QCD performed by several authors
have shown, that depending on the chemical potential
$\mu$ and the temperature $T$ the system displays three possible
phases: 1) chiral symmetry breaking without diquark condensation;
2) mixed phase with nonzero values of both chiral and diquark
condensates; 3) diquark condensation without symmetry
breaking. The system may be described by the thermodynamic
potential $\Omega(\phi, \Delta ; \mu, T)$, where $\phi$ and
$\Delta$ are order parameters, related to the chiral and diquark
condensates respectively (explicit definitions will be given
below). The potential $\Omega$ is expressed in terms of
quantum effective action $\Gamma$ as $\Omega = \Gamma T/V_3$.
Suppose, that the potential $\Omega$ has been calculated
within the framework of some NJL--type model,
i.e. with the gluon sector excluded (except for contribution
of high-frequency modes, giving the necessary attraction).
Consider
for simplicity the phase 3) of the system with $\phi =0, \Delta \neq 0$
and $T=0$. Let $\Omega(\phi=0, \Delta_0; \mu, 0)$ be the stationary
value of the thermodynamic potential, where $\Delta_0$ is the
solution of the gap equation $\partial\Omega/\partial\Delta = 0$
(see below). 
Now we superimpose the nonperturative vacuum gluon fields
on the above picture.
The detailed knowledge of the nonabelian
Meissner effect is unfortunately absent.
Anyway, it is obvious that the corresponding microscopic  
picture is far from being trivial.
The nonlinear character of the 
equations of motion for gluon fields is of prime importance here, 
while usual Meissner effect
for abelian fields is essentially linear phenomenon.
On the contrary, general symmetry arguments tell us, that 
part of gluon degrees of freedom becomes massive if
the colour gauge invariance is spontaneously broken.
It means effective screening of low--frequency modes  
and therefore it is reasonable to assume,
that the formation of the colour gauge invariance --
breaking diquark
condensate should 
lead to the decrease of the gluon 
condensate by some factor,
which we assume to be about a few units
(but, presumably, not to exactly zero value,
as it happens in abelian case).
Then it will be energetically favourable for the
 system to remain in the colour superconducting 
state with $\Delta\neq 0$ only if the 
quantity
\be
\epsilon({\kappa}) = -\left(1-\frac{1}{\kappa}
\right)\>\frac{\beta({\alpha}_s)}{16{\alpha}_s}\;\lll G_{\mu\nu}^a
G_{\mu\nu}^a\rrr  
\label{e10}
\ee
is less than $ \Omega(\phi=\Delta=0; \mu, 0) - 
\Omega(\phi=0, {\Delta}_0; \mu, 0)$.
The factor $\kappa$ in (\ref{e10}) 
represents the unknown rate of decrease of the gluon condensate
due to superconducting diquark state formation.
Note, that $
\epsilon(\kappa\to\infty) = -
\epsilon_{vac}$.
In what follows we will show, that these
two energy gaps have the same order of magnitude
unless the $\kappa$ is close to 1.

It should be noted, that the analogy with the BCS superconductor
in external field is somewhat loose here for at least two 
reasons.
First, strong nonperturbative gluonic fields
are inherent for the QCD vacuum. 
The consistent way of analysis should include a set of 
gap equations for the free energy of the system
depending on gluon and different types of quark condensates,
determining energetically best values for all of 
them simultaneously.
The second point is the relation between scales, characterizing
colour superconductivity and nonperturbative gluon
fluctuations. In particular, only modes with the wave lengths
larger than the Cooper pair radius are responsible for the 
supercurrent while the rest do not admit simple interpretation
in terms of Ginzburg-Landau theory.
We leave the analysis of these complicated
problems for the future.

Another important remark is in order. It might be naively
assumed, that if the system under study is in the 
deconfinement region, the vacuum gluonic content may be
taken as purely perturbative. There are several 
reasons, however, why it is not the case. The most important one
is the following. Finite density breaks Euclidean $O(4)$ rotational
invariance and hence chromoelectric and chromomagnetic components of the   
correlator $\lll {\alpha}_s G^2 {\rrr}_{\mu\neq 0}$ enter on the different
footing. In particular, deconfinement, i.e. zero string tension
is associated with the vanishing of the electric components, while
it is energetically favourable for the magnetic ones to stay nonzero
(the same phenomenon takes place for the temperature phase transition
\cite{s2}).
At the same time it is just strong magnetic field, which is
 able to 
destroy the superconductivity.

Needless to say, that the colour superconductivity 
phenomenon is essentialy nonperturbative
effect since it requires the formation
of the gap, which, being proportional
to ${\Lambda}_{QCD}$, cannot be obtained in 
perturbation theory.

\section{General formalism}

We start with the QCD Euclidean partition function
\be
Z = \int {\cal D} A 
 {\cal D} \bar\psi 
{\cal D} \psi\; exp({-S}) 
\label{e1}
\ee
where
\be
S= \frac14 \int F_{\mu\nu}^a F_{\mu\nu}^a d^4x
+ \int \bar\psi (-i{\gamma}_{\mu}D_{\mu} -im + i\mu {\gamma}_{4})
\psi\> d^4x
\ee
We supress colour and flavour indicies and also introduce
chemical potential $\mu$ (only
the case $N_{f} =2\>;\>N_c =3$ is considered
in this paper).
Performing integration over the gauge fields one gets effective
fermion action in terms of cluster expansion
\be
Z =  
\int {\cal D} \bar\psi 
{\cal D} \psi\; exp\left({-\int d^4x\> L_0  - S_{eff}}\right) 
\label{e3}
\ee
with  $L_0 = 
 \bar\psi (-i{\gamma}_{\mu}{\partial}_{\mu} -im + i\mu {\gamma}_{4})
\psi $
and effective action
$ S_{eff} = \sum\limits_{n=2}^{\infty}
\frac{1}{n!}\> \lll\lll {\theta}^n \rrr\rrr $
where
$
\theta = \int d^4x\> {\bar\psi}(x) g{\gamma}_{\mu}A_{\mu}^a(x)
t^a \psi(x)
$
and double brackets denote irreducible cumulants.

To proceed, one is to make considerable simplification
 of eq.(\ref{e3}). First, 
only the lowest, four-quark interaction term 
is usually kept
in $S_{eff}$.
Second, it is instructive to consider instead of 
the original nonlocal kernel
some idealized local one, respecting the given set of 
symmetries. 
In the problem under study it is common
to choose
either instanton-induced four-fermion vertex 
(the choice, adopted in \cite{rssv,cd}) or 
different versions of the NJL model \cite{awr1,awr2,br} including
the one, motivated by one gluon exchange \cite{awr2}.
In the latter case one gets
\be
S_{eff}
= 
\int d^4x d^4y\> 
 ({\bar\psi}(x) {\gamma}_{\mu}
t^a \psi(x))({\bar\psi}(y) {\gamma}_{\mu}
t^a \psi(y)) D(x-y)
\ee
with $D(x-y)=g^2 \delta^{(4)}(x-y)$ and where the coupling constant $g^2$ 
with the dimension $m^{-2}$ is
 introduced. 
We assume, as it has already been mentioned, that 
this localized form of the kernel does not mimic
all gluonic content of the original theory, in other
words only some part of gluon degrees of freedom 
participates in the condensate formation.
One way to analyse the role, played by other ones,
would be to
consider more realistic nonlocal functions
$D(x-y)$ (which, in principle, encode all necessary
information if we keep only four--quark interaction).
This will be done
elsewhere, while in the present paper we work with
the local form of the action.   
Performing colour,
flavour and Lorentz Fierz \cite{bbr} transformations
and keeping only scalar terms in both $\bar\psi\psi$
and $\psi\psi$ channels, we arrive at
\be
L_{eff}
= g^2 
 \left[({\bar\psi}(x) \Lambda
 \psi(x))({\bar\psi}(y) \Lambda
 \psi(y)) -
 ({\bar\psi}(x) \Phi_{\alpha}
 {\psi^C}(y) )(\bar{{\psi}^C}(y) \Phi_{\alpha}
 \psi(x))\right] 
\label{e6}
\ee
where
$$
\Lambda = \frac{i}{\sqrt{6}}\> {\hat 1}_{c}\>
{\tau}_{F} \; , \;\;  \Phi_{\alpha} 
= \frac{1}{\sqrt{12}}\> {\epsilon}_{\alpha\beta\gamma}
\>{\gamma}_{5}\> {\tau}_{F}^{(2)}
$$ 
and ${\psi}^C = C {\bar\psi}^{T} = {\gamma}_2{\gamma}_4 
\>{\bar\psi}^{T}$.
We note, that with only scalar terms kept, the Lagrangian
(\ref{e6}) is no more chiral invariant.
The attraction in scalar colour antitriplet channel
(which also exists if one starts from the 
instanton--induced interaction) could lead to the 
formation of the condensate, breaking colour $SU(3)$. 
In close anology with \cite{br} we replace the common
coupling constant $g^2$ by two independent constants
$g_1^2$ and $g_2^2$ corresponding to the two 
terms in (\ref{e6}).
Next step is to write down the partition function and to perform
its bosonization. We adopt the standard 
Hubbard--Stratonovich trick and get
$$
Z = 
\int 
 {\cal D} \bar\psi 
{\cal D} \psi\; exp({-\int d^4x\>(L_0 + L_{eff})})=
$$
$$
 = \int {\cal D} \phi {\cal D}\Delta
{\cal D} {\Delta}^{\dagger} exp\left(\>\int d^4 x
\left\{ - \left[\frac{\phi^2}{4g_1^2}
+ \frac{\Delta{\Delta}^{\dagger}}{g_2^2}\right]\right.\right.
$$
\be
+ {\mbox{Tr Ln}}\left.\left.\left(
\begin{array}{cc}
2\Phi C \Delta & i\hat\partial + i(m+\phi) - i{\gamma}_4\mu \\
i{\hat\partial}^{T} - i(m+\phi) + i{\gamma}_4\mu &
2{\Delta}^{\dagger} C^{-1} {\Phi}^{\dagger} \\ 
\end{array}
\right)\right\}\right)
\ee

For the system of the massless quarks at the phase 3) and $T=0$
the thermodynamic potential reads (we remind, that $Z\sim exp(-S)
\sim exp(+\Gamma)$)
\be
\Omega = \frac{\Gamma}{V_4} = \frac{|\Delta|^2}{g_2^2} -
\frac{1}{V_4}
{\mbox{Tr Ln}}\left(
\begin{array}{cc}
2\Phi C \Delta &
i\hat\partial - i{\gamma}_4\mu
\\
i{\hat\partial}^{T} + i{\gamma}_4\mu &
2{\Delta}^{\dagger} C^{-1} {\Phi}^{\dagger} \\
\end{array}
\right)
\label{e17}
\ee
The value of the diquark condensate is determined by the 
gap equation
\be
\left.\frac{\partial\Omega}{\partial\Delta}\right|_{\Delta = {\Delta}_0} =0
\label{gap}
\ee
By going in (\ref{e17}) to wave number--frequency space
and making use of the gap equation (\ref{gap}), we arrive 
at the following expression for the thermodynamic 
potential at its minimum
\be
 \Omega(\phi=\Delta=0; \mu, 0) - 
\Omega(\phi=0, {\Delta}_0; \mu, 0) \simeq
\frac{{\Delta}_0^2}{g_2^2\cdot ln\left(\frac{M^2}{{\Delta}_0^2}
\right)}
\label{e100}
\ee
where $M$ is the NJL cutoff which is typically about 
$0.8\> Gev$ \cite{awr1,br}. 
Alternatively, the cutoff may enter via the formfactor 
or the instanton zero-mode. In the NJL--type calculations the 
values of the cutoff $M$ and the coupling constant $g_2^2$ are fitted
simultaneously, but no unique "standard" fit exists so far \cite{kik}.
To estimate the r.h.s. of (\ref{e100})
we have taken for cutoff $M=0.8\> Gev$, for coupling 
$g^2 = 12g_2^2 = (15\div 40)\> {Gev}^{-2} $ and the value
of the gap in the diquark scalar sector $\Delta_0 =
(0.1 \div 0.15)\> Gev$. With these parameters one gets
\be
 \Omega(\phi=\Delta=0; \mu, 0) - 
\Omega(\phi=0, {\Delta}_0; \mu, 0) \sim
(1\div 5)\cdot 10^{-3}\> {Gev}^4
\label{e121}
\ee
To be on a robust quantitative footing and to consider finite
temperatures one can replace the estimate (\ref{e121})
by the result of direct numerical calculations of the thermodynamic
potential performed in \cite{br}. Our result (\ref{e121}) is
larger than the corresponding value, presented in \cite{br}. 
The discrepancy may be due to different value of the coupling 
constant $g^2$ adopted in \cite{br}. Needless to say, that the
larger is the estimate of (\ref{e100}) the larger 
is the critical field extinguishing superconductivity.

Now let us estimate the expression (\ref{e10}).
We use different sets of data from \cite{dd} and take for
the gluon condensate 
$$G_2 = 
\lll \frac{{\alpha}_s}{\pi}\>G_{\mu\nu}^a
G_{\mu\nu}^a\rrr = (0.014\div 0.026)\> Gev^4
$$ 
Then for two flavours one gets in one loop
$$
\epsilon(\kappa) = \left(1-\frac{1}{\kappa}\right)
\cdot \left( 11 - \frac43 \right)\cdot\left( \frac{1}{32}
\right)\;\lll \frac{{\alpha}_s}{\pi}\>G_{\mu\nu}^a
G_{\mu\nu}^a\rrr \sim 
$$
\be
\sim 
\left(1-\frac{1}{\kappa}\right)\cdot (4\div 8)\cdot 10^{-3}\> {Gev}^{4}
\label{e19}
\ee
It is seen, that (\ref{e100}) and (\ref{e19}) have the same 
order of magnitude for $\kappa\ge 2$.

It should be noted, that the estimate (\ref{e121})
given above is rather optimistic in the following sense.
If one naively assume the colour superconductor to be the
BCS one and take its typical parameters, for example, from \cite{rssv},
then one has at $T=0$ for the 
value of the critical magnetic field 
\be
H_{crit}^2 = 0.64\cdot (\Delta_0)^2 \cdot (m_q p_F)
\label{w2}
\ee
where $p_F^2 = \mu^2 - m_q^2$, $m_q$ is the constituent 
quark mass.
We take $\mu = (0.4\div 0.5)\>Gev$ and 
corresponding
$\Delta_0 = (0.04\div 0.10)\>Gev$ from the paper
\cite{rssv}
which are smaller, than the typical values we 
have analysed before.
The maximum of (\ref{w2})
with respect to $p_F$ at the fixed $\mu$
is reached for $\mu = \sqrt{2} p_F$ and it gives  
\be
H_{crit}^2 = (0.8\div  8 )\cdot 10^{-4}\>Gev^4
\label{w3}
\ee
Strictly speaking we are not allowed to apply
formulas like (\ref{w2}) to the colour fields and
interpret them in terms of Meissner effect, 
but we note, that (\ref{w3}) even without any
numerical factors
is
about order of magnitude smaller than (\ref{e19}).

\section{Finite density effects}

The comparison made in the previous section
was intuitively based on the analogy with the 
Meissner effect in ordinary superconductors. 
The actual value of the gluon
condensate in (\ref{e19}) was taken to be the vacuum 
one. 
This is not quite correct, however.
Even without the formation of any diquark condensate, the
gluon condensate in the hadronic matter is different
from that in the vacuum.
In order to get an idea about such dependence,
let us consider effective dilaton Lagrangian \cite{ms}
\be
L(\sigma) = \frac12 ({\partial}_{\mu} \sigma)^2 - V(\sigma)
\;\; ;\;\;\; V(\sigma) = 
\frac{\lambda}{4}
\> {\sigma}^4 \left( \ln\frac{\sigma}{\sigma_0} -\frac14\right)
\ee
where dilaton field is defined according to 
$$
\frac{m_0^4}{64|\epsilon_{v}|}\> \sigma^4(x) = 
 - \theta_{\mu\mu}(x) = -\frac{\beta({\alpha}_s)}{4{\alpha}_s}\;
G_{\mu\nu}^a (x)
G_{\mu\nu}^a (x)  
$$
and 
$$
\lambda = \frac{m_0^4}{16|\epsilon_{v}|} \;\; ;\;\;\; 
{\sigma}_0^2 = \frac{16|\epsilon_{v}|}{m_0^2}
$$
where $\epsilon_{v}= \frac14 \>\lll \theta_{\mu\mu}\rrr$ 
is the nonperturbative vacuum energy 
density and $m_0$ - dilaton (i.e. glueball $0^{++}$ in our case) 
mass.
Low energy dilaton physics can be used for description
of the gluon condensate behaviour at finite density and
temperature \cite{a}.
In the chiral limit masses of nucleons in QCD
are determined by the nucleon-dilaton 
vertex  $L_{int} = m_{N}^* \bar q q$ with the 
effective mass $m_{N}^* = m_{N}\cdot(\sigma/\sigma_0)$.
In isotopically symmetric system the energy density takes
the form:
\be
\epsilon(\sigma_{n}, n) = V(\sigma_n) - V(\sigma_0) +
2\int\limits_0^{p_F} \frac{d^3p}{(2\pi)^3}\> \sqrt{p^2 + {m_N^*}^2} 
\ee
where $n=n_n + n_p = 2p_F^3/3{\pi}^2$ is the baryon density and
the chemical potential $\mu^2 = p_F^2 + {m_N^*}^2$.
Being interested in the densities close to the nuclear density
$n_0^{1/3} = 0.1\>Gev$ we minimize the total energy density
with respect to the dilaton field 
$$
\left.
\frac{\partial \epsilon}{\partial \sigma}\right|_{\sigma = \sigma_n} = 0
$$
and find 
in the region $n m_N \ll 16|\epsilon_{v}|$
\be
\lll \frac{{\alpha}_s}{\pi}\>G_{\mu\nu}^a
G_{\mu\nu}^a\rrr_{n} = 
\lll \frac{{\alpha}_s}{\pi}\>G_{\mu\nu}^a
G_{\mu\nu}^a\rrr_{0}\cdot\left(1-c\>\frac{n}{n_0}\right)
\label{y3}
\ee
where the factor $c=m_N n_0/4|\epsilon_v| \simeq 0.05$. It means, that 
the decrease of the condensate at the densities 
of interest 
is about $(10\div 25)\%$.

However, we should point out, that the matter may have
already  the structure
of cold quark-gluon
plasma instead of being hadronic at the densities of interest.
Presumably it would require another type of estimates for the 
value of the gluon condensate.
The reader is referred to the paper \cite{son}
where it is argued, that the long--wave part of colour magnetic 
field which can lead to the Meissner effect is supressed
compared to the critical field
by the perturbative coupling constant (which is assumed to be
small due to asymptotic freedom). 
It is important to distinguish the quantum fields
considered in \cite{son} which  
propagate in the quark-gluon plasma from the nonperturbative 
fields discussed by us here, hence the magnitude of the latter
is an "external parameter" with respect to the colour
superconductivity problem itself.
This question needs further investigation.

\section{Discussion}

We have confined ourselves in this letter
to a very modest aim -- to compare the energy gap,
typical for the colour superconductor and
the contribution to the vacuum energy density, coming from 
the gluon condensate. We argued, that {\it if} there is a competition
between colour antitriplet scalar diquark and gluon
condensates, which is natural to assume from
abelian analogy, {\it then} it happens on
the same energy scale and therefore to take 
this effect into account is important for 
the selfconsistent picture.
It is clear,
that further, more quantitative
 analysis is needed.      

The idealized case of two flavours was investigated
in the paper. For chemical potentials exceeding the strange 
quark mass the $N_f=N_c=3$ scenario is more 
physical.
It seems, that 
inclusion of the third flavour
does not change our analysis crucially. Moreover,
due to the phenomenon of colour-flavour "locking" \cite{awr2},
eight of the nine $((N_c^2 -1)_{SU(3)} + 1_{U(1)})$ 
gauge degrees of freedom acquires mass due to the Higgs mechanism
in $N_f =3$ case
and therefore colour superconductivity is
"complete". According to the line of reasoning adopted in this
paper, it means, that $\kappa({N_f=3})$ should be
larger than $\kappa({N_f=2})$ (for the scalar channel).

To conclude, we have discussed the role played by gluon
degrees of freedom in the colour superconductivity
and, in particular, we have argued, that nonperturbative 
gluon fluctuations might be strong enough to destroy
the phenomenon. More quantitative analysis of this problem
is in progress now.

\bigskip

{\bf Acknowledgements}

The work was supported in part by 
by RFFI-DFG grant 96-02-00088G.
The work of B.K. was also supported by RFFI grant
97-02-16406. V.Sh. acknowledges
the ICFPM-INTAS 96-0457 grant.
The authors are grateful to 
I.~Dobroskok, D.~Ebert, J.~Ho\v{s}ek, H.-J.~Pirner,
K.~Rajagopal and Yu.A.~Simonov 
for useful discussions and comments. 
V.Sh. would like to thank theory group
of {\v{R}}e\v{z} Nuclear
Physics Institute and especially
J.~Ho\v{s}ek for kind hospitality.


\begin{thebibliography}{99}
\bibitem{awr1}
M.Alford, K.Rajagopal, F.Wilczek, Phys.Lett.
{\bf B422} (1998) 247.
\bibitem{rssv}
R.Rapp, T.Schaefer, E.Shuryak, M.Velkovsky,
Phys.Rev.Lett. {\bf 81} (1998) 53.
\bibitem{br}
J.Berges, K.Rajagopal,
Nucl.Phys. {\bf B538} (1999) 215.
\bibitem{awr2}
M.Alford, K.Rajagopal, F.Wilczek,
Nucl.Phys. {\bf B537} (1999) 443.
\bibitem{cd}
G.~Carter, D.~Diakonov, 
Nucl.Phys. {\bf A642} (1998) 78.
\bibitem{bbr}
J.~Bijnens, C~Bruno, E. de Rafael, Nucl.Phys. {\bf B390}
(1993) 501;\\
D.~Ebert, H.~Reinhardt, M.~Volkov, Progr.Part.Nucl.Phys. 
{\bf 33} (1994) 1.  
\bibitem{ev}
D.~Ebert, M.~Volkov, Phys.Lett. {\bf B272} (1991) 86.
\bibitem{s}
Yu.~Simonov, Yad.Fiz., {\bf 57} (1994) 1491.
\bibitem{s2}
E.~Gubankova, Yu.~Simonov, Phys.Lett. {\bf B360}
(1995) 93.
\bibitem{kik}
S.~Klevansky, Rev.Mod.Phys. {\bf 64} (1992) 649;\\
U.~Vogl, W.~Weise, Progr.Part.Nucl.Phys. {\bf 27} (1992) 195;\\
G.~Ripka, {\it Quarks Bound by Chiral Fields}, Clarendon Press,
Oxford, 1997.
\bibitem{dd}
H.G.~Dosch, Proc. of the Int. School of
Physics "Enrico Fermi", "{\it Selected
topics in nonperturbative QCD}", IOS Press, 1996;\\
M.D'Elia, A.~DiGiacomo, E.~Meggiolaro,
Phys.Lett. {\bf B408} (1997) 315;\\
S.~Narison, Phys.Lett. {\bf B387} (1996) 162.
\bibitem{ms}
A.A.~Migdal, M.A.~Shifman, Phys.Lett. {\bf B114} (1982) 445;\\
J.~Ellis, J.~Lanik, Phys.Lett. {\bf B150} (1985) 289.
\bibitem{a}
N.O.~Agasyan, JETP Lett. {\bf 57} (1993) 208;\\
N.O.~Agasyan, D.~Ebert, E.~Ilgenfritz, Nucl.Phys. {\bf A637}
(1998) 135.
\bibitem{son}
D.T.~Son, hep-ph/9812287.
\end{thebibliography}
\end{document}